\documentstyle[amssymb,aps,12pt]{revtex}
\begin{document}
\draft
\author{Sergio De Filippo\cite{byline}}
\address{Dipartimento di Fisica ''E. R. Caianiello'', Universit\`{a} di Salerno\\
Via Allende I-84081 Baronissi (SA) ITALY\\
Tel: +39 089 965 229, Fax: +39 089 965 275, e-mail: defilippo@sa.infn.it\\
and \\
Unit\`{a} INFM Salerno}
\date{\today}
\title{Gravity-induced entropy in the quantum motion of a macroscopic body.}
\maketitle

\begin{abstract}
It is shown that a recently proposed model for the gravitational interaction
in non relativistic quantum mechanics may turn to be relevant to the
derivation of the second law of thermodynamics. In particular, the spreading
of the probability density of the center of mass of an isolated macroscopic
body does not imply delocalization of the wave function, but on the contrary
it corresponds to an entropy growth.
\end{abstract}


The monotone time dependence of the entropy of an isolated system and the
emergence of the arrow of time have long been staying as a debated issue,
since the birth of classical statistical mechanics. A brief sketch of some
contributions to that long-standing debate within quantum physics\cite
{vonneumann,landau,schroedinger,pauli,lindblad,zurek} can be found in Ref. 
\cite{gemmer}, where a quantum approach to the derivation of the second law
of thermodynamics was proposed. In it a non unitary dynamics of a gas-system
ensues from the peculiarity of the model, where the interaction with the
container obeys a suitable form of generalized microcanonicity, namely its
only effect is to produce entanglement, without affecting energy
conservation.

In this letter we want to explore the possibility that even for a genuinely
isolated system an entropy growth takes place, within a non relativistic
quantum description, owing to the gravitational interaction. In order to do
that we use a model for the gravitational interaction in non relativistic
quantum mechanics\cite{defilippo}, where self-interactions are treated on
the same footing as mutual interactions between different bodies. This is
achieved by duplicating the ordinary degrees of freedom, with the
introduction of a (red) partner for every ordinary (green) particle and of a
Newton gravitational interaction between (meta)particles of different color
only, and restricting the (meta)state space by a suitable constraint. Once
red metaparticles are traced out, the resulting non unitary dynamics
reproduces both the classical aspects of the gravitational interaction and
wave function localization of macroscopic bodies\cite{defilippo}.

Even in a one particle model like the Schroedinger-Newton theory, without
added unobservable degrees of freedom, one can exhibit stationary localized
states \cite{kumar}. However, apart from the price paid in abandoning the
traditional linear setting of QM, they sound quite unrealistic, as the
initial linear momentum uncertainty is expected to lead to a spreading of
the probability density in space. On the other hand a theory of wave
function localization has to keep localization during time evolution. Our
model offers a way out of this apparent paradox, leading to a spreading that
consists in the emergence of delocalized ensembles of localized pure states.

To be specific, let $H[\psi ^{\dagger },\psi ]$ denote, following Ref. \cite
{defilippo}, the second quantized non-relativistic Hamiltonian of a finite
number of particle species, like electrons, nuclei, ions, atoms and/or
molecules, according to the energy scale. For notational simplicity $\psi
^{\dagger },\psi $ denote the whole set $\psi _{j}^{\dagger }(x),\psi
_{j}(x) $ of creation-annihilation operators, i.e. one couple per particle
species and spin component. This Hamiltonian includes the usual
electromagnetic interactions accounted for in atomic and molecular physics.
To incorporate gravitational interactions including self-interactions, we
introduce complementary creation-annihilation operators $\chi _{j}^{\dagger
}(x),\chi _{j}(x)$ and the overall Hamiltonian 
\begin{equation}
H_{G}=H[\psi ^{\dagger },\psi ]+H[\chi ^{\dagger },\chi
]-G\sum_{j,k}m_{j}m_{k}\int dxdy\frac{\psi _{j}^{\dagger }(x)\psi
_{j}(x)\chi _{k}^{\dagger }(y)\chi _{k}(y)}{|x-y|},
\end{equation}
acting on the tensor product $F_{\psi }\otimes F_{\chi }$ of the Fock spaces
of the $\psi $ and $\chi $ operators, where $m_{i}$ denotes the mass of the $%
i$-th particle species and $G$ is the gravitational constant. While the $%
\chi $ operators are taken to obey the same statistics as the original
operators $\psi $, we take advantage of the arbitrariness pertaining to
distinct operators and, for simplicity, we choose them commuting with one
another: $[\psi ,\chi ]$ $_{-}=[\psi ,\chi ^{\dagger }]_{-}=0$.

The metaparticle state space $S$ is identified with the subspace of $F_{\psi
}\otimes F_{\chi }$ including the metastates obtained from the vacuum $%
\left| 0\right\rangle =\left| 0\right\rangle _{\psi }\otimes \left|
0\right\rangle _{\chi }$ by applying operators built in terms of the
products $\psi _{j}^{\dagger }(x)\chi _{j}^{\dagger }(y)$ and symmetrical
with respect to the interchange $\psi ^{\dagger }\leftrightarrow \chi
^{\dagger }$, which, as a consequence, have the same number of $\psi $
(green) and $\chi $ (red) metaparticles of each species. In particular the
most general metastate corresponding to one $j$-particle states is
represented by 
\begin{equation}
\left| \left| f\right\rangle \right\rangle =\int dx\int dyf(x,y)\psi
_{j}^{\dagger }(x)\chi _{j}^{\dagger }(y)\left| 0\right\rangle
,\;\;f(x,y)=f(y,x),
\end{equation}
with one green and one red $j$-metaparticle. This is a consistent definition
since the overall Hamiltonian is such that the corresponding time evolution
is a group of (unitary) endomorphisms of $S$. If we prepare a pure $n$%
-particle state, represented in the original setting - excluding
gravitational interactions - by 
\begin{equation}
\left| g\right\rangle \doteq \int d^{n}xg(x_{1},x_{2},...,x_{n})\psi
_{j_{1}}^{\dagger }(x_{1})\psi _{j_{2}}^{\dagger }(x_{2})...\psi
_{j_{n}}^{\dagger }(x_{n})\left| 0\right\rangle ,
\end{equation}
its representation in $S$ is given by the metastate 
\begin{equation}
\left| \left| g\otimes g\right\rangle \right\rangle =\int
d^{n}xd^{n}yg(x_{1},...,x_{n})g(y_{1},...,y_{n})\psi _{j_{1}}^{\dagger
}(x_{1})...\psi _{j_{n}}^{\dagger }(x_{n})\chi _{j_{1}}^{\dagger
}(y_{1})...\chi _{j_{n}}^{\dagger }(y_{n})\left| 0\right\rangle .
\label{initial}
\end{equation}
As for the physical algebra, it is identified with the operator algebra of
say the green metaworld. In view of this, expectation values can be
evaluated by preliminarily tracing out the $\chi $ operators and then taking
the average in accordance with the traditional setting.

While we are talking trivialities as to an initial metastate like in Eq. (%
\ref{initial}), that is not the case in the course of time, since the
gravitational interaction in the overall Hamiltonian produces entanglement
between the two metaworlds, leading, once $\chi $ operators are traced out,
to mixed states of the physical algebra. It was shown in Ref. \cite
{defilippo} that the ensuing non-unitary evolution induces both an effective
interaction mimicking gravitation, and wave function localization.

It was shown also that, omitting the internal wave function, a localized
metastate of an isolated homogeneous spherical macroscopic body of radius $R$
and mass $M$ can be represented by 
\begin{equation}
\tilde{\Psi}_{0}(X,Y)\propto \exp \frac{-\left| X-Y\right| ^{2}}{2\Lambda
^{2}}\exp \frac{-\left| X+Y\right| ^{2}}{2\Lambda ^{2}},\;\;\Lambda ^{2}=%
\frac{\hslash }{\sqrt{\alpha GM^{3}/R^{3}}},  \label{gaussian}
\end{equation}
where $\alpha \sim 10^{0}$ and $G$ is the gravitational constant. In Eq. (%
\ref{gaussian}) $X$ and $Y$ are respectively the position of the center of
mass of the green and the red metabody. The first factor is proportional to
the wave function of the relative motion and, for bodies of ordinary density 
$\sim 1gm/cm^{3}$ and whose mass exceeds $\sim 10^{11}$ proton masses, it
represents its ground state\cite{defilippo}. The second factor is
proportional to the wave function of the center of metamass $(X+Y)/2$ and
spreads in time as usual for the free motion of the center of mass of a body
of mass $2M$, so that after a time $t$, in the absence of external forces,
the metawave function becomes 
\begin{equation}
\tilde{\Psi}_{t}(X,Y)\propto \exp \frac{-\left| X-Y\right| ^{2}}{2\Lambda
^{2}}\exp \frac{-\left| X+Y\right| ^{2}/4}{\Lambda ^{2}/2+i\hslash t/M}%
\equiv \exp \left[ -\alpha _{0}\left| X-Y\right| ^{2}\right] \exp \left[
-\alpha _{t}\left| X+Y\right| ^{2}\right] .  \label{evolvedgaussian}
\end{equation}
In order that this may be compatible with the assumption that gravity
continuously forces localization, the spreading of the physical state must
be the outcome of a growth of the corresponding entropy. This initially
vanishes, since the initial metawave function is unentangled: 
\begin{equation}
\tilde{\Psi}_{0}(X,Y)\propto \exp \frac{-X^{2}}{\Lambda ^{2}}\exp \frac{%
-Y^{2}}{\Lambda ^{2}},
\end{equation}
and then the corresponding physical state obtained by tracing out $Y$ is
pure. If one evaluates the physical state according to 
\begin{equation}
\rho _{t}(X,X^{\prime })=\int dY\tilde{\Psi}_{t}(X,Y)\tilde{\Psi}_{t}^{\ast
}(X^{\prime },Y),
\end{equation}
one finds that the space probability density is given by 
\begin{equation}
\rho _{t}(X,X)=\left[ \frac{8\alpha _{0}(\alpha _{t}+\bar{\alpha}_{t})}{\pi
(\alpha _{t}+\bar{\alpha}_{t}+2\alpha _{0})}\right] ^{3/2}\exp \left[ -\frac{%
8\alpha _{0}(\alpha _{t}+\bar{\alpha}_{t})}{(\alpha _{t}+\bar{\alpha}%
_{t}+2\alpha _{0})}X^{2}\right] \propto \exp \frac{-2\Lambda ^{2}X^{2}}{%
\Lambda ^{4}+2\hslash ^{2}t^{2}/M^{2}}  \label{probability}
\end{equation}
Parenthetically it is worth while to remark that this spreading of the
probability density is slower than the one ensuing from the spreading of the
wave function in the absence of the gravitational self-interaction, which
leads to 
\begin{equation}
\rho _{t}(X,X)\propto \exp \frac{-2\Lambda ^{2}X^{2}}{\Lambda ^{4}+4\hslash
^{2}t^{2}/M^{2}},  \label{traditionalprobability}
\end{equation}
and that both are extremely slow, as their typical time, for macroscopic
bodies of ordinary density, is $\sim 10^{3}\sec $ independently from the
mass, as can be checked by means of Eqs. (\ref{probability}) and (\ref
{traditionalprobability}).

If this spreading is due to entropy growth only, rather than to the usual
spreading of the wave function, the corresponding entropy $S_{t}$ is
expected to depend approximately on the ratio between the final and the
initial space volumes roughly occupied by the two gaussian densities,
according to 
\begin{equation}
S_{t}\sim K_{B}\frac{3}{2}\ln \left[ \frac{\alpha _{t}+\bar{\alpha}%
_{t}+2\alpha _{0}}{2(\alpha _{t}+\bar{\alpha}_{t})}\right] ,
\label{approxentropy}
\end{equation}
at least for large enough times. (Linear momentum probability density does
not depend on time.) Of course this corresponds to approximating the mixed
state by means of an ensemble of $N$ equiprobable localized states, which is
legitimate if $N$ is large enough. In order to evaluate the entropy of the
state represented by $\rho _{t}(X,X^{\prime })$ and to check Eq. (\ref
{approxentropy}), we use the possibility, in this approximation, of linking
the entropy 
\begin{equation}
S_{t}=-K_{B}\text{ }Tr\left[ \rho _{t}\ln \rho _{t}\right] =K_{B}\ln N
\label{equientropy}
\end{equation}
with the purity 
\begin{equation}
Tr\left[ \rho _{t}^{2}\right] =\frac{1}{N},  \label{equipurity}
\end{equation}
where of course 
\begin{equation}
\rho _{t}^{2}(X,X^{\prime })=\int dX^{\prime \prime }\rho _{t}(X,X^{\prime
\prime })\rho _{t}(X^{\prime \prime },X^{\prime }).
\end{equation}
By an explicit computation we get 
\begin{equation}
Tr\left[ \rho _{t}^{2}\right] =\int dX\rho _{t}^{2}(X,X)=\frac{\left[
4\alpha _{0}(\alpha _{t}+\bar{\alpha}_{t})\right] ^{3}}{\left[ \left(
2\alpha _{t}\bar{\alpha}_{t}+6\alpha _{t}\alpha _{0}+6\bar{\alpha}_{t}\alpha
_{0}+2\alpha _{0}^{2}\right) ^{2}-4\left( \bar{\alpha}_{t}-\alpha
_{0}\right) ^{2}\left( \alpha _{t}-\alpha _{0}\right) ^{2}\right] ^{3/2}},
\end{equation}
and, for large times, namely small $\alpha _{t}$, one can keep in this
result just the leading term in $\alpha _{t}$, that is 
\begin{equation}
Tr\left[ \rho _{t}^{2}\right] \sim \left( \frac{\alpha _{t}+\bar{\alpha}_{t}%
}{2\alpha _{0}}\right) ^{3/2},
\end{equation}
which, by using Eqs. (\ref{equientropy},\ref{equipurity}), gives 
\begin{equation}
S_{t}\sim -K_{B}\frac{3}{2}\ln \left( \frac{\alpha _{t}+\bar{\alpha}_{t}}{%
2\alpha _{0}}\right) =K_{B}\frac{3}{2}\ln \left( \frac{\Lambda ^{4}+4\hslash
^{2}t^{2}/M^{2}}{\Lambda ^{4}}\right) ,
\end{equation}
which differs from the leading term in Eq.(\ref{approxentropy}) by an
irrelevant quantity $(3/2)K_{B}\ln 2$.

It is worth while to remark that, while the present model is expected to be
just a low energy approximation to a conceivable more general theory, its
present application to a free motion is expected basically to reproduce the
possible exact outcome of the latter. In fact the analysis refers to the
rest frame of the probability density and implies exceedingly small
velocities, as can be checked by taking the Fourier transform of the wave
function in Eq. (\ref{evolvedgaussian}).

This result is an encouraging hint towards the possibility of deriving the
second law of thermodynamics for genuinely isolated systems from what most
people call quantum gravity, or say from quantum mechanics with a proper
inclusion of gravity. Of course much remains to be done both with reference
to the consideration of \ further instances, physically more comprehensive
than just free motions of macroscopic bodies, and to the extension of the
present approach, to cope with energies higher than the ones where the non
relativistic model is appropriate.

Finally it should be made clear that the present letter addresses what can
be called fundamental entropy growth, which in principle could be defined
even for the universe as a whole. For real, only approximately isolated,
systems one would expect that usually, however small the coupling with the
environment may be, the corresponding entanglement entropy\cite{gemmer} is
easily large enough to overshadow the fundamental one. This is the
thermodynamical counterpart of the overshadowing of fundamental decoherence 
\cite{defilippo} by the environment induced one\cite{zurek1}. Just as for
this latter issue one may expect that, while waiting for a possible future
experimental detection, fundamental decoherence may play a role with
reference to the measurement problem in QM, likewise fundamental entropy
growth is a natural candidate to address the issue of the arrow of time.
While the entropy growth we are talking about is extremely small, that is
only natural for a possible breaking of time reversal symmetry responsible
for the time arrow. This smallness, on the other hand, is connected with the
high localization threshold at $\sim 10^{11}$ proton masses, amply
compatible with the present lower bounds at $\sim 10^{3}$\ proton masses\cite
{rae,arndt}.

To sum up, for a generic isolated system the present picture leads to
surmise that time flow is characterized by the monotone increase of the
entanglement between observable and unobservable degrees of freedom. While
this view can be taken in principle within more conventional attempts to
build a quantum theory of gravity as well\cite{ellis}, a peculiar feature of
the present approach is the simple and unambiguous definition of the
unobservable degrees of freedom and of the overall dynamics, which makes the
model a viable computational tool.

\newpage Acknowledgments - Financial support from M.U.R.S.T., Italy and
I.N.F.M., Salerno is acknowledged

\end{document}